# Raman Signatures of Ferroic Domain Walls captured by Principal Component Analysis

G. F. Nataf[1,2,3,*], N. Barrett[2], J. Kreisel[1,4], M. Guennou[1]

[1]Materials Research and Technology Department, Luxembourg Institute of Science and Technology, 41 Rue du Brill, L-4422 Belvaux, Luxembourg
[2]SPEC, CEA, CNRS, Université Paris-Saclay, CEA Saclay, 91191 Gif-sur-Yvette Cedex, France
[3]Department of Materials Science, University of Cambridge, 27 Charles Babbage Road, Cambridge CB3 0FS, UK
[4]Physics and Materials Science Research Unit, University of Luxembourg, 41 Rue du Brill, L-4422 Belvaux, Luxembourg
*corresponding author: gn283@cam.ac.uk

**Abstract**

Ferroic domain walls are currently investigated by several state-of-the art techniques in order to get a better understanding of their distinct, functional properties. Here, Principal Component Analysis (PCA) of Raman maps is used to study ferroelectric domain walls (DWs) in $LiNbO_3$ and ferroelastic DWs in $NdGaO_3$. It is shown that PCA allows to quickly and reliably identify small Raman peak variations at ferroelectric DWs and that the value of a peak shift can be deduced - accurately and without *a-priori* - from a first order Taylor expansion of the spectra. The ability of PCA to separate the contribution of ferroelastic domains and DWs to Raman spectra is emphasized. More generally, our results provide a novel route for the statistical analysis of any property mapped across a DW.

Ever since it has been realized that domain walls (DWs) in ferroic materials can present novel structural properties that do not exist in the bulk domains, researchers have considered a new device paradigm where the DWs, rather than the domains, are the active element. This field has been coined "Domain Boundary engineering" [1,2] or "Domain wall nanoelectronics" [3]. The exploitation of the DWs nanometric size as well as their different functional properties presents a high potential for industrial innovation [2,3]. Such functionalities include ferroelectric/polar DWs in ferroelastic $CaTiO_3$ [4–6] or $SrTiO_3$ [7], electric conductivity of domain walls in the insulating $BiFeO_3$ [8–10] or $Pb(Zr,Ti)O_3$ [11–13], and photo-induced conductivity of DWs in lithium niobate ($LiNbO_3$, LN) single crystals [14,15]. The example of local conductivity in the highly insulating and transparent LN is particularly appealing.

LN itself has been the subject of intensive studies for many years due to its ferroelectric, piezoelectric, pyroelectric, acoustic, electro-optical or photorefractive properties. It is suitable for applications in photorefractive devices, holographic memories, frequency doublers, etc [16]. At room temperature, it has a rhombohedral structure with space group *R3c*. Planes of oxygen atoms are arranged in a distorted hexagonal close-packed arrangement, with the interstices alternately filled by Li and Nb atoms. In the ferroelectric phase, Li and Nb atoms are slightly displaced from their position of high symmetry, defining the direction of the spontaneous polarization. The ferroelectric transition in LN is not ferroelastic – as opposed to many classical ferroelectrics - which makes LN a model of the kind. Only two domain states are possible, both with polarization along the rhombohedral *c*-axis, and therefore only 180°-domain walls are observed. Interesting anomalies are observed in the vicinity of these DWs: unexpected optical contrast [17], strain on length scales of micrometers [18], anomalies in the Raman [19–25] and dielectric [26] spectra and photo-induced conduction [14,15]. These observations are usually discussed in term of variations in defect concentrations but the precise underlying mechanisms remain unclear and are debated in literature [19–21]. In the case of photo-induced conduction, an artefact effect due to the electrodes cannot be excluded.

Here, we present a study of DWs with Raman spectroscopy, which is a non-destructive, non-contact, and defect-sensitive method for investigating ferroic materials. Generally speaking, the changes of the Raman spectra in the vicinity of DWs are small, with typical peak shifts below 1 $cm^{-1}$. Thus, a large number of spectra and good statistics are usually needed for obtaining reliable results of the mostly very subtle changes. A further difficulty is that the typical ferroelectric or ferroelastic DW width is less than 10 nm [27], much less than the spatial resolution used in Raman spectroscopy. As a result, the Raman signal of interest is superimposed on a high background signal coming from the adjacent domains. The results are commonly analysed by peak fitting [21,23–25] or by subtracting a spectrum at the domain wall from a spectrum far from it [20,22] when Raman modes overlap, making fitting difficult. In this paper we demonstrate that a statistical analysis, namely Principal Component Analysis (PCA) can overcome such difficulties and provides a rapid and robust way for the analysis of Raman data across DWs. Our work is inspired from literature reports on the use of PCA for Raman mapping of different chemical species [28,29], to correct peak shifts smaller than the spectral resolution in Raman spectroscopy [30,31], or even to detect phase transition under pressure [32]. PCA has also been very useful in combination with a number of other experimental techniques: quantitative analysis of X-ray photoelectron spectroscopy imaging [33], scanning transmission electron microscopy image analysis [34], energy loss electron spectroscopy [35], etc. Here, we describe the interest of PCA for a quick and semi-quantitative analysis of DW signatures.

In the first part of this article we present simulations showing that PCA can be used to rapidly detect peak shifts, peak width and peak intensity variations in Raman spectra. In the second part, we apply PCA to the case of 180°-DWs in LN where contrast is only expected at the wall and compare the results with a standard fitting procedure. In the third part, we extend our discussion to variations of Raman modes at ferroelastic DWs in $NdGaO_3$ where adjacent domains are also expected to contribute to contrast.

1. Identification and localization of changes by PCA

1.1. Theory

Raman measurements at domain walls usually consist of point-by-point mapping across a domain wall. The resulting maps typically contain hundreds of spectra. PCA is used to express the data in such a way as to highlight similarities and differences in a large set of spectral data. This can be used to identify specific spectral features otherwise hidden by the noise, or automatically identify characteristic signatures in a Raman map. These features are called principal components (PCs) [36]. Each spectrum $S$ obtained from the dataset, here a 3D matrix corresponding to a map of Raman spectra, can be described as a linear combination of PCs:

$$S = S_a + \sum_k t_k \cdot PC_k \qquad (1)$$

where $S_a$ is the mean spectrum calculated as the normalized sum of all the spectra and $t_k$ is a scalar, called the *score*, representing the weight of each component in the spectrum $S$. The number of PCs is equal to the number of spectra in the dataset.

These PCs are obtained from the singular value decomposition of the spectra in the mapping dataset $X$. Let us assume that $X$ is centred, *i.e.* column means have been subtracted and are now equal to zero. $X$ is expressed as the product of three matrices:

$$X = U \cdot D \cdot W^t \qquad (2)$$

where the columns of $U$ are the left-singular vectors of $X$, the diagonal entries of the diagonal matrix $D$ are the singular values of $X$ and the column of $W$ are the right-singular vectors of $X$ ($W^t$ is the conjugate transpose matrix of $W$). In the vocabulary of the PCA, the columns of $W$ are the PCs and the projections of the data on the PCs are the scores $t_k$ given by:

$$X \cdot W = U \cdot D \cdot W^t \cdot W = U \cdot D \qquad (3)$$

The first PC ($PC_1$) is required to have the largest possible variance, in order to explain the most significant changes in the spectra. The second PC ($PC_2$) is computed under the constraint of being orthogonal to the first PC and to have the largest possible variance. The other components are computed likewise [37].

The next stage of the analysis consists then in determining the number of components needed to describe the data: if too many PCs are retained, one might try to give a physical meaning to noise; if too few PCs are retained, one might miss essential information. A key parameter to determine the number of PCs is the proportion of total variance accounted for by each PC. This proportion $P$ is:

$$P(PC_k) = \frac{V(PC_k)}{\sum V(S)} \qquad (4)$$

where $V(S)$ and $V(PC_k)$ are the variances of each spectrum and each PC, respectively. The lower is $P$ the less information is contained in the PC. However, defining a threshold value for the number of PCs to be retained is arbitrary [28,29]: the examples in Fig. 2 show that a PC which accounts for only 0.9% of the total variance can still have a physical meaning.

In Raman micro-spectroscopy, PCA has been commonly used for chemometric analysis. It allows to gain both spectral and spatial information when imaging various chemical species distributed over several micrometers [28,29]. It has also been used to follow phase transitions [32]. In these cases, species or phases had well defined signatures and PCA produces components that can be directly interpreted as spectral signatures of chemical species. Conversely, in the case of 180°-DWs, the changes of the Raman spectra expected in the vicinity of the DW are small [19], leading to small alterations of the Raman spectra with respect to an average spectrum corresponding to the neighbouring domains. Thus the interpretation of the components is less straightforward.

### 1.2. Simulated DW Raman signatures

To understand subtle changes and their signatures close to DWs, we perform simulations of Raman spectra with *Python 2.7*. We consider a series of 300 spectra containing each a single Lorentzian peak centred at a frequency of 750 cm$^{-1}$, with a FWHM of 20 cm$^{-1}$ and an intensity $I$ of 100. The discrete spectral resolution is of 1 cm$^{-1}$. The spectrum number is associated to a spatial position $X$ (in μm) along a line, *i.e.* one spectrum per micron. In order to determine the physical signification of these deviations, we classify them as peak shifts, full width at half maximum (FWHM) changes and peak intensity variations. The objective of this part is to identify the characteristic shapes of the PCs associated to these spectrum perturbations.

In the first simulation, a DW is simulated by shifting the Raman peak in the central part of the line by -0.5 cm$^{-1}$ with respect to the two neighbouring "domains", as shown in Fig. 1a. PCA is applied to the series of simulated spectra: Fig. 1d shows the mean spectrum and the only PC obtained. This PC resembles the first derivative of the Lorentzian peak. Indeed, adding the derivative to the Lorentzian peak will increase the intensity on the left side of the peak maximum and decrease the intensity on the right side, shifting the Lorentzian to lower frequencies.

In the second simulation series, the FWHM of twenty Lorentzians is decreased by 2.5% before PCA is applied, as shown in Fig. 1b. Figure 1e shows the PC obtained: it is symmetric with an increase of intensity at the centre. Adding the PC to the Lorentzian peak will decrease the intensity on its shoulders, reducing its FWHM.

We consider in the third simulation a series of spectra containing two Lorentzian peaks with the same intensity, centred at 400 cm$^{-1}$ and 1200 cm$^{-1}$. The intensity of the Lorentzian peak at higher frequency is progressively decreased by 0.5%, decreasing the intensity ratio between the second and the first Lorentzian, as shown in Fig. 1c. Figure 1f shows the mean spectrum and the PC obtained which resemble a Lorentzian peak with a negative intensity.

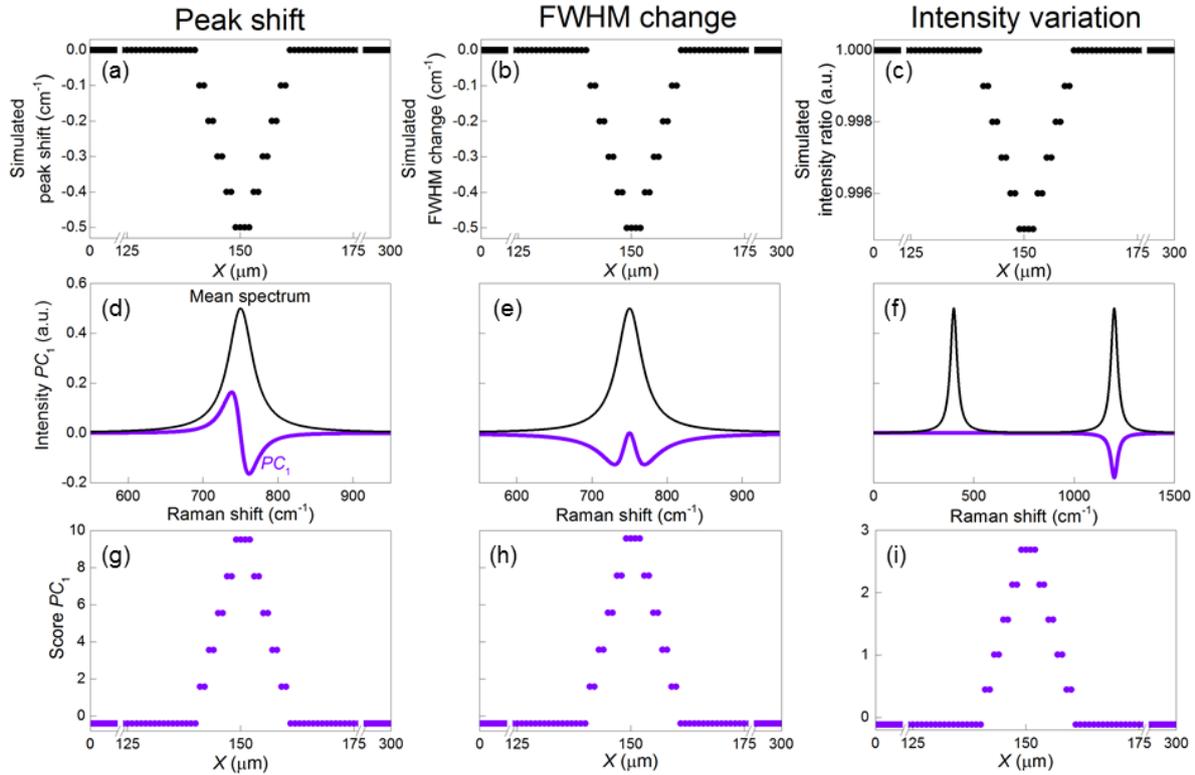

**Figure 1. PCA results for simulated hypothetical spectra. Three hypothetical domain wall signatures are applied to 20 Lorentzian: (a) a shift of a single peak, (b) a change in the FWHM of a single peak and (c) an intensity variation between two peaks. The resulting signature of the principal component and mean spectrum are displayed in parts (d-f), while parts (g-i) present the corresponding scores for every spectrum.**

Figures 1g-i are the scores of the PCs. They reveal the spatial position of the twenty spectra affected by the changes. Qualitatively, the product of the PC and its score gives the "trend" of the changes. As an example, the negative peak shift is characterized by an increase of the intensity of the PC on the left side of the Lorentzian maximum and a decrease on its right side – because the PC score is positive.

### 1.3. Influence of noise

Because PCA is a statistical treatment, it is expected to be affected by the signal-to-noise ratio of the original dataset. We study the influence of noise in the following simulation. Twenty spectra are subjected to a shift of +0.5 cm$^{-1}$. A white noise with a normal distribution – centre at 0 and standard deviation $\sigma$ - is added to every spectrum. We define the signal to noise ratio as $SNR = I/\sigma$, where $I$ denotes the intensity of the Lorentzian Raman peak.

As shown in Fig. 2a, with $SNR = 1000$, $PC_1$ still exhibits the expected characteristic shape and the score clearly shows the spatial position of the shifted spectra (Fig. 2d). With $SNR = 100$, $PC_1$ is noisy and its score does not allow to define a precise value of the shift (Fig. 2b,e). With $SNR = 10$, PCA is no longer able to detect the frequency shift (Fig. 2c) and the score reveals only the white noise (Fig. 2f). The absolute values of the scores increase when the SNR decreases because of the higher variations in the intensity of the spectra.

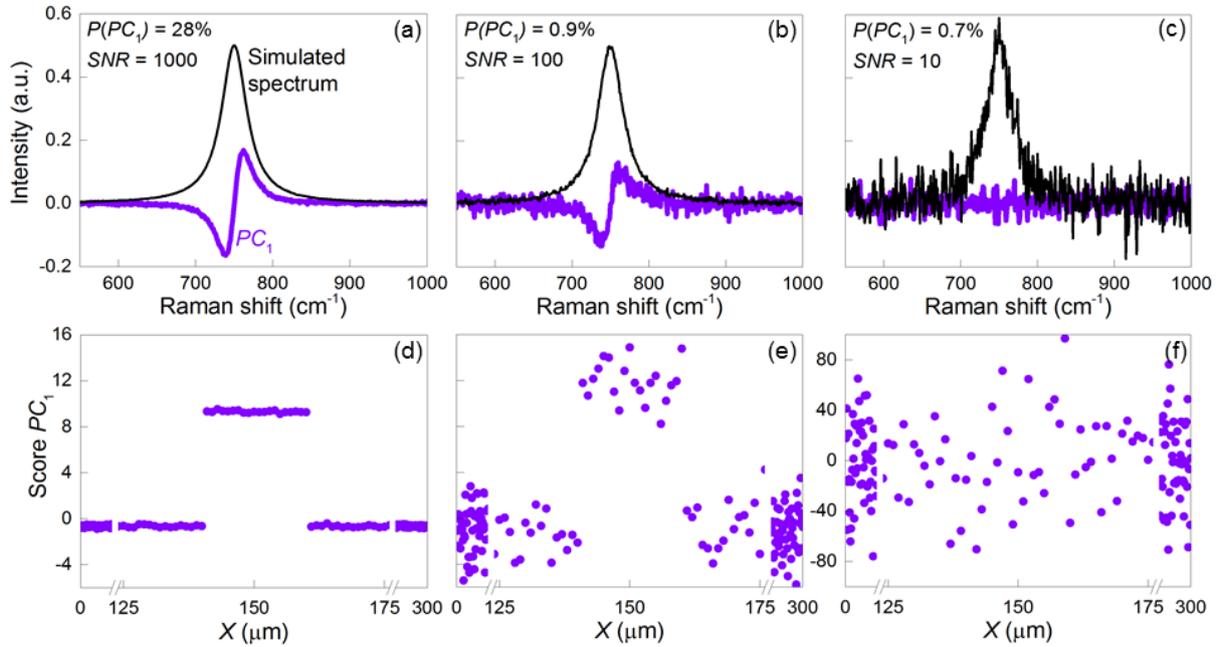

**Figure 2. Influence of noise.** $PC_1$ obtained when performing PCA on a peak shift of +0.5 cm$^{-1}$ with a SNR of (a) 1000, (b) 100 and (c) 10. (d-f) are the corresponding scores.

The ability of PCA to detect subtle changes depends on their magnitude. Figure 3 shows how it is possible to detect a frequency shift, even with $SNR = 10$, by increasing the magnitude of the shift. When twenty Lorentzian are shifted by 3 cm$^{-1}$, $PC_1$ resembles again the first derivative of the Lorentzian peak (Fig. 3a) but the score is noisy and does not allow a precise location of the shifted spectra (Fig. 3d). The contrast in the scores increases for frequency shift of 5 cm$^{-1}$ (Fig. 3e) and 10 cm$^{-1}$ (Fig. 3f). In other words, the SNR imposes a limit on the minimal peak shift that can be detected.

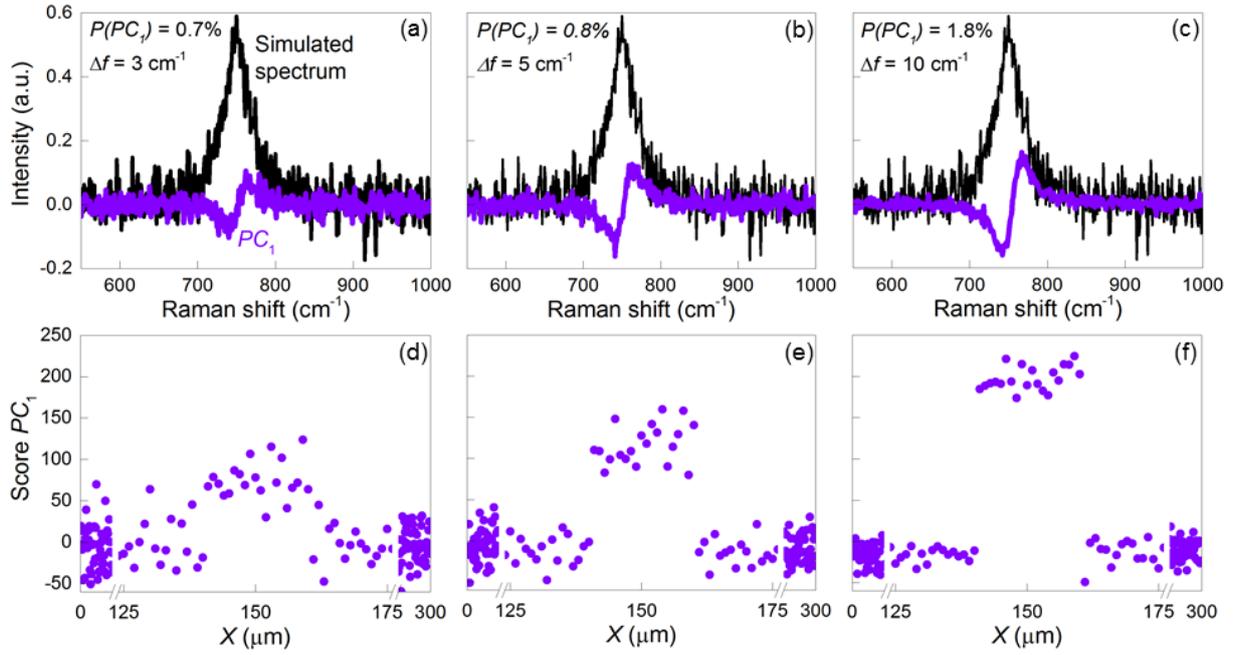

**Figure 3. Influence of the amplitude of the frequency shift on noisy spectra. PC obtained when performing PCA on spectra with a SNR = 10 and a frequency shift of (a) 3 cm$^{-1}$, (b) 5 cm$^{-1}$ and (c) 10 cm$^{-1}$. (d-f) are the corresponding scores.**

If the same white noise is added to every spectrum, PCA is able to distinguish it from the Lorentzian peak and to give an accurate value of the frequency shift (not shown). Any recurrent noise in a set of Raman spectra is therefore easily identified and filtered with PCA.

### 1.4. Ability to identify several simultaneous changes

Raman modes often undergo a peak shift and a FWHM change at the same time (for example when doping LN [19]). If the frequency and the FWHM of a Lorentzian peak are simultaneously modified, PCA still gives only one PC. In the example of Fig. 4, a peak shift of +0.5 cm$^{-1}$ is applied combined with a decrease of the FWHM by 2.5%. As shown in Fig. 4a, this PC resembles a linear combination of the PCs obtained in Fig. 1d and Fig. 1e. Figure 4b shows the score of the PC: the twenty modified Lorentzian peaks exhibit the same value. In other words, a simultaneous change of frequency and peak width is characterized by an asymmetric PC with maxima of opposite signs.

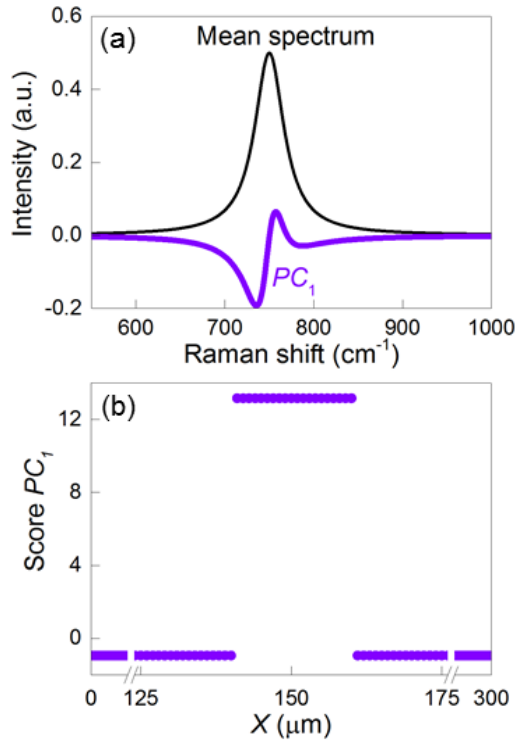

**Figure 4. PCA on a simultaneous peak shift and FHWM variation. (a) Lorentzian peak unmodified and $PC_1$ (b) Scores of $PC_1$.**

2. **Investigation of 180°-DWs: example of LiNbO$_3$**

Ferroelectric DWs in lithium niobate with congruent [19,20,22–24] or stoichiometric [20,22,25] compositions have been extensively studied by Raman spectroscopy [19,20,22–25]. The usual measurement consists of point-by-point mapping across a DW. One reason for this interest is that 180°-domains by symmetry must have the same Raman spectrum. Any contrast observed near the DW can then be related to the DW itself. Peak shifts [19,22,25] and peak intensity variations [20,21,23–25] of the Raman modes have been observed but their relation to structure remains unclear. They are usually explained by defect concentration differences, or stress induced field variations.

We investigated the z-face of a 5 mol% magnesium doped lithium niobate single crystal. We performed a map of 2x8 µm², in 0.1 µm steps, as shown in the inset of Fig. 5d. Each spectrum was acquired in 0.5 s, with an exciting laser line at 442 nm and a spot size of 0.65 µm. After acquisition of the Raman map we used PCA to identify the main changes in the spectra near domain walls.

Figures 5a-c show three PCs and Fig. 5d-f the corresponding scores. In Fig. 5a-c, the mean spectrum, corresponding to the sum of all the spectra in the data set, is also plotted for reference and is considered as the typical Raman spectrum of LN. The number of PCs studied is limited to three because they account for 85% of the total variance. The fourth component - which account for only 0.04% - and all the others, contain only noise.

In Fig. 5a, $PC_1$ is similar to the mean Raman spectrum. By direct comparison with the results of the simulation, this PC describes changes of intensity of the spectrum. These variations are usually referred to as 'noise' in Raman experiments: they are due to fluctuations of the laser intensity and drift of the sample holder during the measurement. Thus, the score of this PC is not at all correlated with the domain structure (Fig. 5d).

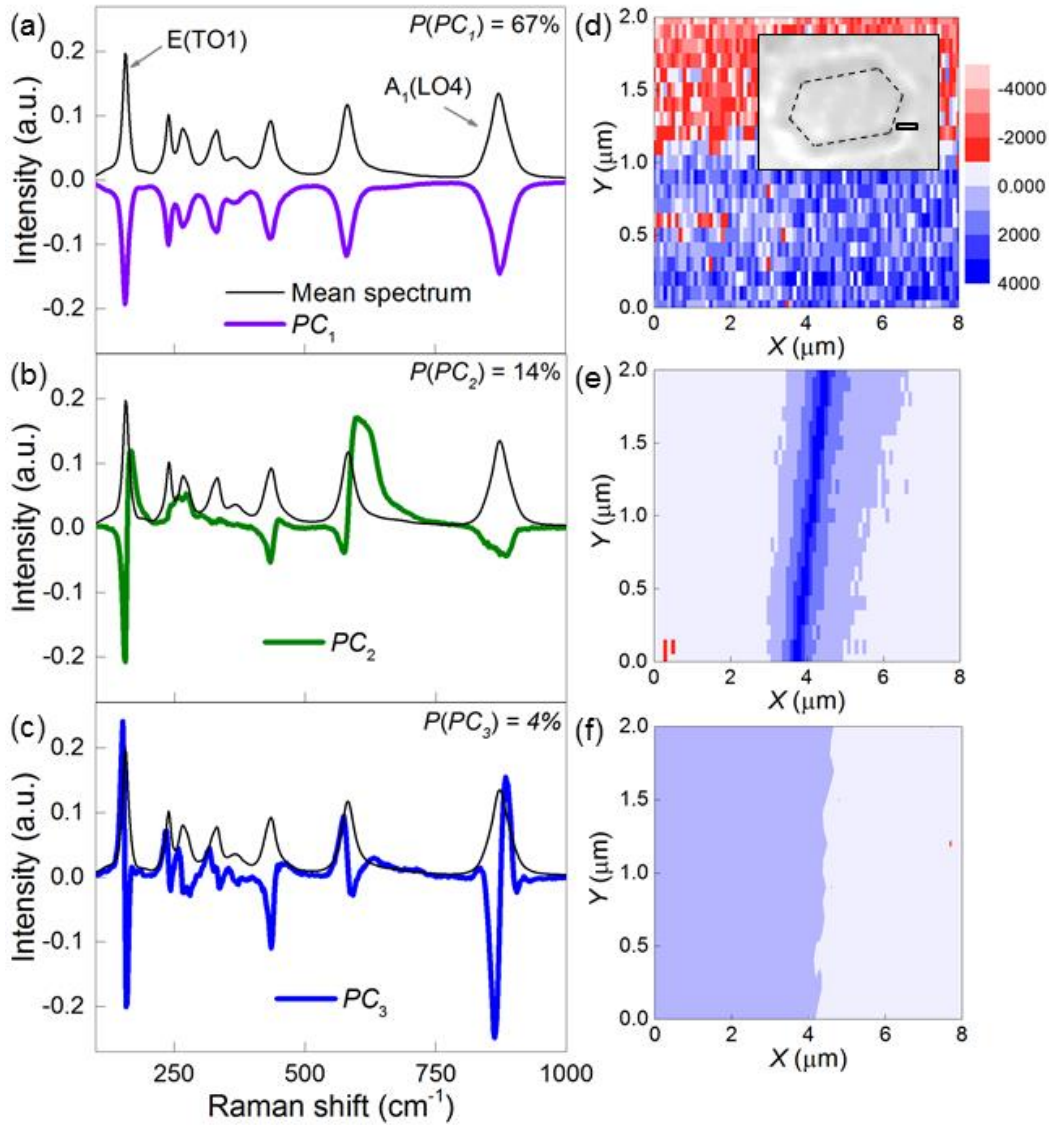

**Figure 5. PCA Raman maps of ferroelectric domains and domain walls in LiNbO$_3$. Mean spectrum and (a) first PC, (b) second PC, (c) third PC. (d-f) are the corresponding scores. The inset in (d) is an optical image of the domain with the investigated area indicated by the black rectangle and the DWs by dashed lines below them.**

Figures 5b,c present features centred at the same frequencies as the different Raman modes: they are similar to what was observed in the simulation of peak shifts or peak intensity variations. At 154 cm$^{-1}$, $PC_2$ resembles the first derivative of a Lorentzian and by comparison with simulations, is interpreted as a shift of the E(TO1) mode centred at 154 cm$^{-1}$. By comparison with simulations, the large change of intensity of $PC_2$ near 615 cm$^{-1}$ is a growing peak. $PC_3$ is dominated by two features: a shift of E(TO1) at 154 cm$^{-1}$ and a shift of A$_1$(LO4)

at 870 cm$^{-1}$. Figures 5d-f show the score of the first, second and third PCs, respectively. The score of $PC_2$ presents a high contrast in the position of the domain wall as determined by optical microscopy. Interestingly, the score of $PC_3$ reveals a small contrast between domains which, by symmetry, is not expected. It is in fact due to the internal field of LN as explained below.

The interpretation of the PCA as performed on the full spectrum is that (i) the E(TO1) peak position is slightly shifted between up- and down-polarized domains and shifts significantly at the domain wall (ii) the $A_1$(LO4) peak position is different in domains. In order to study these two modes individually, we performed other PCAs whereby we restricted the frequency range to include only the peak of interest: between 40 and 200 cm$^{-1}$ for E(TO1), between 750 and 1000 cm$^{-1}$ for $A_1$(LO4). The first two PCs are shown in Fig. 6a for E(TO1) and Fig. 6b for $A_1$(LO4). They are similar to the features observed in Fig. 5. For both modes, $PC_1$ resembles the Raman mode and $PC_2$ is similar to the derivative. The PCA shows that the sum of $PC_1$ and $PC_2$ describes well the E(TO1) and $A_1$(LO4) modes. Since $PC_1$ is similar to the mean signal, we can consider that $PC_2$ alone describes the frequency shift.

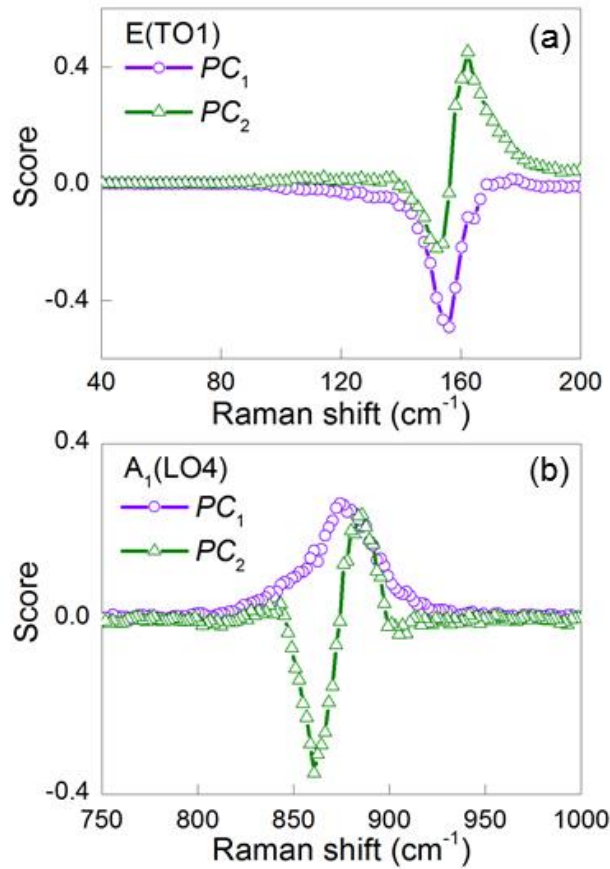

**Figure 6. Results of the PCA analysis on individual modes. Scores of the first and second components for (a) E(TO1) and (b) $A_1$(LO4).**

In order to compare the results of the PCA with the results of the standard fitting procedure, all spectra have been fitted with Voigt functions using the classical analysis approach. Figure 7a compares the normalized score of $PC_2$ with the normalized value of the E(TO1) peak shift. The behaviour of $PC_2$ and the value of the peak shift are in good agreement. Fig. 7b shows the same comparison for $A_1(LO4)$ mode. Again, a good agreement is observed.

By considering the true Raman spectrum as the sum of the mean spectrum and its derivative the spectra can be expanded in a Taylor series to first order:

$$S(f) = b_1 \cdot L(f) + b_2 \cdot \frac{dL}{df} \qquad (5)$$

where $b_1$ and $b_2$ are the Taylor coefficients, $L$ is the lineshape function and $\frac{dL}{df}$ is the derivative of the spectrum with respect to the frequency $f$. Within the hypothesis of Lorentzian (or Voigt) lineshape:

$$b_1 = A \qquad \text{and} \qquad b_2 = A \cdot \Delta f \qquad (6)$$

where $A$ is the peak maximum of the spectrum and $\Delta f$ is the peak shift. Thus, once a reference spectrum is chosen and its first derivative calculated, the Taylor coefficients can be estimated by linear regression of the reference spectrum and its derivative with respect to the complete series of spectra. The calculated regression coefficients are used to determine the peak shift $\Delta f$.

The full algorithm to determine the peak shift is the following:
(1) The mean spectrum is chosen as the reference spectrum and is normalized.
(2) Its first derivative is calculated numerically.
(3) Taylor coefficients are estimated by multiple linear regression. In matrix notation:

$$B = (X^t \cdot X)^{-1} \cdot X^t \cdot S \qquad (7)$$

where the two columns of $X$ are the reference spectrum and its derivative, the columns of $S$ are the measured spectra pixel by pixel, the two columns of $B$ are the Taylor coefficients ($b_1$, $b_2$).
(4) The frequency shifts are determined by:

$$\Delta f = \frac{b_2}{b_1} \qquad (8)$$

At the domain wall, a shift of ~0.25 cm$^{-1}$ is observed in E(TO1) (Fig. 7c). It extends over ~1.5 μm, which is about twice the theoretical spot size (0.65 μm), i.e. more than one would expect from a typical 10 nm DW. While it is conceivable that modification of physical properties occurs on a larger scale than 10 nm (like strain fields near DWs in LN), there are also many experimental factors that can lead to a degradation of the instrumental response (slight defocusing, tiny misalignment of the sample, etc.), and it is difficult to make a conclusive statement about the true spatial extension of the effects. Furthermore, a shift of ~0.1 cm$^{-1}$ is observed between the domains. The behaviour of the calculated (PCA and Taylor expansion) and fitted (standard procedure) peak shifts are in good agreement: we obtain a slope of 1 with a Pearson's correlation coefficient of 0.99 when plotting one as a function of the other. Figure 7d shows the same comparison for $A_1(LO4)$ mode. A negative shift of ~0.3 cm$^{-1}$ is observed when going from the left to the right domain. The behaviour of the calculated and fitted peak shifts are again in good agreement (slope 1, Pearson's coefficient 0.99).

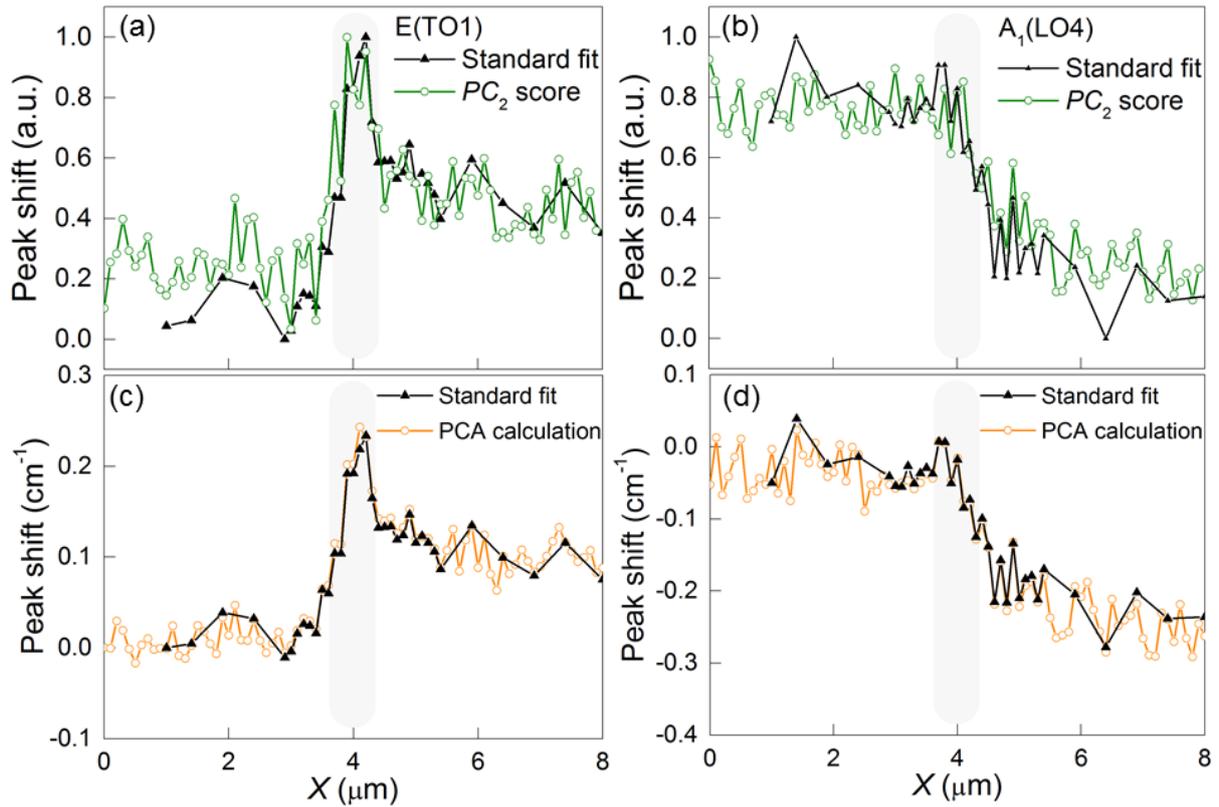

**Figure 7. Peak shifts deduced from calculations and fitting procedures. Normalized components for (a) E(TO1) and (b) $A_1$(LO4). Absolute values for (c) E(TO1) and (d) $A_1$(LO4).**

A previous systematic study of the influence of doping on the Raman signature at domain walls showed that (i) the frequency shift at the domain wall involves a combination of effects resulting from the defect structure and the electric (strain) field in the vicinity of the wall (ii) the Raman contrast that we observe between domains can be explained by the internal field induced by polar defect structures [19].

3. **Ferroelastic domain walls: example of NdGaO$_3$**

Ferroelastic domains have by definition different crystallographic orientations that come with different orientations of the Raman tensor, which creates a contrast in Raman spectroscopy between two adjacent domains. This allows Raman mapping of ferroelastic domains, as is it customarily done [38]. Because of the large size of our laser spot (~0.65 µm) compared with the typical DW width (< 10 nm), the Raman signal at the boundary between two ferroelastic domains contains a weighted average of the different Raman signal from both domains, possibly masking an extra, specific, signature of the DW. Identifying a specific signature of the DW is therefore much more challenging than in case of a pure ferroelectric DW.

In principle, the Raman contrast between the two domains cancels out for specific orientation and scattering geometries. This is a direct consequence of the strain compatibility condition obeyed by typical domains in bulk crystals. In practice though, it may be very difficult to identify and select this geometry. Besides, the cancellation would rely on a perfect realization of Raman selection rules, which are in reality only approximately fulfilled because of misorientations, optics, etc.

Here, we illustrate how PCA can be used to extract information from a Raman mapping across a ferroelastic domain wall. We investigate neodymium gallium oxide ($NdGaO_3$), which has an orthorhombic structure corresponding to the space group *Pbnm*, the most common structure among perovskites at room temperature [39], leading to 24 Raman modes among which 18 are well-resolved [40]. Its phonon-dispersion curves and phonon eigenvectors have already been calculated [40]. The spontaneous strain coefficients, determined from the lattice parameters [41], are $e_{11} = 8.2 \times 10^{-4}$ and $e_{12} = 6.5 \times 10^{-3}$.

The Raman mapping of a $NdGaO_3$ single crystal was carried out with a helium-neon laser at 633 nm. Domains were observed with an optical microscope working with polarized light in order to select DWs almost orthogonal to the surface. We performed a map of 4x16 µm$^2$, in 0.1 µm steps, as shown in the inset of Fig. 8c.

As shown in Fig. 8c, the score of $PC_1$ is constant in the first domain, decreases on ~3 µm and reaches a new constant value in the second domain. The two scores have opposite sign. Thus, $PC_1$ describes the domain structure and the weighted average of the Raman signal from both domains. In Fig. 8a, $PC_1$ is compared with two spectra extracted from the domains. $PC_1$ shows intensity variations: subtracting $PC_1$ from a spectrum of domain 1 gives a spectrum characteristic of domain 2. $PC_2$ must then describe an extra contribution observed in the vicinity of the wall, as evidenced by its score (Fig. 8d). In Fig. 8b, $PC_2$ is compared with a spectrum from domain 1. It exhibits mainly intensity variations of the Raman modes.

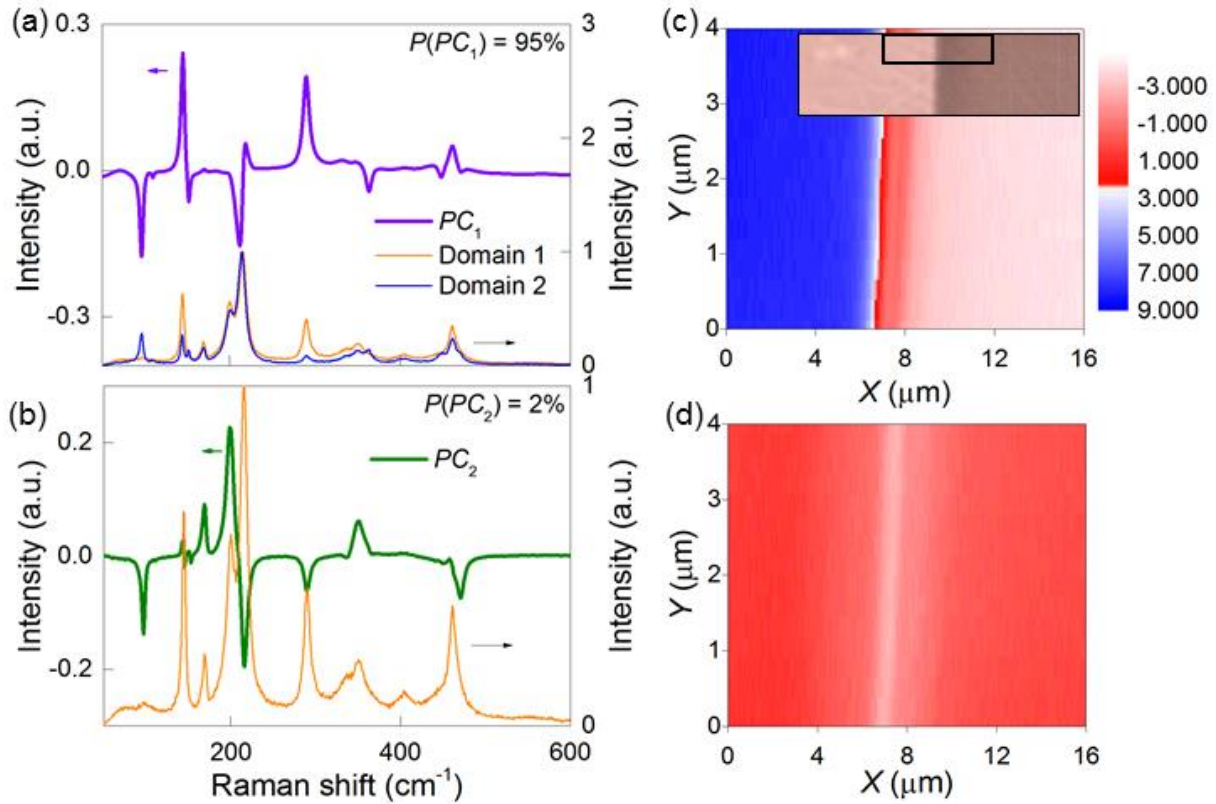

**Figure 8. PCA Raman maps of ferroelastic domain and domain walls in NdGaO$_3$.** (a) Comparison of $PC_1$ and two spectra from domain 1 and domain 2. (b) Comparison of $PC_2$ with a spectrum from domain 1. (c) Scores of $PC_1$ for every spectrum. (d) Scores of $PC_2$ for every spectrum. The inset in (c) is an optical image of the domains with the investigated area indicated by the black rectangle.

Thus, PCA allows Raman mapping of ferroelastic domains and DWs. Contrary to lithium niobate, where the contrast between domains was described by the *PC* with the smallest variance ($PC_3$), the significant differences in intensities between the spectra of ferroelastic domains are described by $PC_1$, which accounts for 95% of the total variance. Ferroelastic DWs are oriented in such a way as to maintain strain compatibility between two adjacent domains, in order to minimize stress and elastic energy. Therefore, it is not surprising that we do not observe frequency shifts characteristic of strain variations at domain walls, contrary to DWs in lithium niobate where strain on length scales of micrometers have been reported [18]. It is not the purpose of this paper to give a detailed interpretation of the specific signature at the DW and how it relates to changes in structural and physical properties (octahedral tilts, birefringence, etc.); this is currently under investigation and requires more in-depth analyses.

**Conclusion**

In summary, our simulations show that PCA is a powerful tool to identify peak shifts, peak width and peak intensity variations in a 3D matrix corresponding to a map of Raman spectra. Raman measurements performed on 180°-DWs in lithium niobate evidence that PCA can be used to quickly identify and quantitively characterized small variations at DWs. PCA is also a promising tool to isolate the contribution of ferroelastic DWs.

Apart from DWs, Raman micro-spectroscopy and PCA could be used as complementary tools to map with a microscopic resolution the distribution of strain in materials, as required in silicon [42], graphene [43] or multilayer ceramic capacitors [44]. The use of PCA could also be extended to 3D matrix obtained through other experimental techniques, such as surface potential maps [45].

**Acknowledgements**

This work was supported by the Luxembourg National Research Fund (FNR) under project CO-FERMAT FNR/P12/4853155/Kreisel.